\RequirePackage{fancyvrb}

\documentclass[sigconf,authorversion]{acmart}

\pdfoutput=1

\usepackage{amssymb}
\usepackage[edges]{forest}

\apptocmd{\sloppy}{\hbadness 10000\relax}{}{}

\usepackage{anyfontsize}

\DeclareMathOperator*{\argmax}{arg\,max}

\usepackage[caption=false,lofdepth,lotdepth]{subfig}

\usepackage[inline, shortlabels]{enumitem}

\usepackage{colortbl}
\newcolumntype{x}[1]{>{\centering\arraybackslash\hspace{0pt}}m{#1}}

\usepackage{multirow}
\usepackage[binary-units=true]{siunitx}
\usepackage[strict]{csquotes}
\usepackage{tikz}
\usetikzlibrary{arrows.meta}
\usetikzlibrary{backgrounds}
\usetikzlibrary{calc}
\usetikzlibrary{fit}
\usetikzlibrary{patterns}
\usetikzlibrary{positioning}
\usetikzlibrary{shadows}
\usetikzlibrary{shapes}
\usetikzlibrary{trees}
\usepackage{pgfplots}
\pgfplotsset{compat=1.14}

\usepackage[xlatin,abbreviations]{foreign}

\usepackage[acronym,hyperfirst=false]{glossaries}
\glsdisablehyper

\newcommand{\aref}[1]{\hyperref[#1]{Appendix~\ref{#1}}}

\newenvironment{customlegend}[1][]{%
    \begingroup
    \csname pgfplots@init@cleared@structures\endcsname
    \pgfplotsset{#1}%
}{%
    \csname pgfplots@createlegend\endcsname
    \endgroup
}%

\def\addlegendimage{\csname pgfplots@addlegendimage\endcsname}

\makeatletter
\tikzset{%
on layer/.code={%
        \pgfonlayer{#1}\begingroup
        \aftergroup\endpgfonlayer
        \aftergroup\endgroup
    },
-|-/.style={
to path={
        (\tikztostart) -| ($(\tikztostart)!#1!(\tikztotarget)$) |- (\tikztotarget)
        \tikztonodes
    }
},
-|-/.default=0.5,
}
\makeatother

\newcommand{\specialcell}[2][c]{\begin{tabular}[#1]{@{}c@{}}#2\end{tabular}}

\newacronym{BIOS}{BIOS}{Basic Input/Output System}
\newacronym{C2}{C\&C}{Command and Control}
\newacronym{COW}{COW}{Copy-On-Write}
\newacronym{DDOS}{DDoS}{Distributed Denial-of-Service}
\newacronym{DOS}{DOS}{Denial Of Service}
\newacronym{IDS}{IDS}{Intrusion Detection System}
\newacronym{IPC}{IPC}{Inter-Process Communication}
\newacronym{MAC}{MAC}{Mandatory Access Control}
\newacronym{MAEC}{MAEC}{Malware Attribute Enumeration and Characterization}
\newacronym{P2P}{P2P}{Peer-to-peer}
\newacronym{RPC}{RPC}{Remote Procedure Call}
\newacronym[shortplural={OSs}, longplural={Operating Systems}]{OS}{OS}{Operating System}
\newacronym{STIX}{STIX}{Structured Threat Information eXpression}
\newacronym{UEFI}{UEFI}{Unified Extensible Firmware Interface}
\newacronym[shortplural={VLANs}, longplural={Virtual Local Area Networks}]{VLAN}{VLAN}{Virtual Local Area Network}
\newacronym[shortplural={VMs}, longplural={Virtual Machines}]{VM}{VM}{Virtual Machine}
\newacronym{MOO}{MOO}{Multi-Objective Optimization}
\newacronym{SOO}{SOO}{Single-Objective Optimization}

\VerbatimFootnotes

\setcopyright{acmlicensed}

\begin{document}
\pagestyle{plain}
\pagestyle{empty}
\copyrightyear{2019}
\acmYear{2019}
\acmConference[ACSAC '19]{2019 Annual Computer Security Applications Conference}{December 9--13, 2019}{San Juan, PR, USA}
\acmBooktitle{2019 Annual Computer Security Applications Conference (ACSAC '19),
December 9--13, 2019, San Juan, PR, USA}
\acmPrice{15.00}
\acmDOI{10.1145/3359789.3359792}
\acmISBN{978-1-4503-7628-0/19/12}

\title{Survivor: A Fine-Grained Intrusion Response and Recovery Approach for Commodity Operating Systems}

\author{Ronny Chevalier}
\orcid{0000-0002-7479-4988}
\affiliation{\institution{HP Labs}}
\affiliation{\institution{CentraleSup\'elec, Inria, CNRS, IRISA}}
\email{ronny.chevalier@hp.com}

\author{David Plaquin}
\affiliation{\institution{HP Labs}}
\email{david.plaquin@hp.com}

\author{Chris Dalton}
\affiliation{\institution{HP Labs}}
\email{cid@hp.com}

\author{Guillaume Hiet}
\orcid{0000-0002-7176-9760}
\affiliation{\institution{CentraleSup\'elec, Inria, CNRS, IRISA}}
\email{guillaume.hiet@centralesupelec.fr}

\renewcommand{\shortauthors}{Chevalier et al.}

\begin{abstract}
    Despite the deployment of preventive security mechanisms to
    protect the assets and computing platforms of users,
    intrusions eventually occur.
    We propose a novel intrusion survivability approach to withstand ongoing intrusions.
    Our approach relies on an orchestration of fine-grained recovery and
    per-service responses (\eg privileges removal).
    Such an approach may put the system into a degraded mode.
    This degraded mode prevents attackers to reinfect the system or to achieve their
    goals if they managed to reinfect it.
    It maintains the availability of core functions while waiting for patches to be deployed.
    We devised a cost-sensitive response selection process
    to ensure that while the service is in a degraded mode,
    its core functions are still operating.
    We built a Linux-based prototype and evaluated
    the effectiveness of our approach against different types of intrusions.
    The results show that our solution removes the effects of the intrusions,
    that it can select appropriate responses, and that it allows services to survive when reinfected.
    In terms of performance overhead,
    in most cases, we observed a small overhead,
    except in the rare case of services that write many small files asynchronously in a burst, where we
    observed a higher but acceptable overhead.
\end{abstract}



\maketitle
\section{Introduction}
Despite progress in preventive security mechanisms such as
cryptography, secure coding practices, or network security,
given time, an intrusion will eventually occur.
Such a case may happen due to technical reasons
(\eg a misconfiguration, a system not updated, or an unknown vulnerability)
and economic reasons~\cite{kshetri2006simple}
(\eg do the benefits of an intrusion for criminals outweigh their costs?).

To limit the damage done by security incidents,
intrusion recovery systems help administrators restore
a compromised system into a sane state.
Common limitations are that they do not preserve availability~\cite{ashvin2005taser,kim2010retro,hsu2006BFF}
(\eg they force a system shutdown)
or that they neither stop intrusions
from reoccurring nor withstand
reinfections~\cite{xiong2009shelf,ashvin2005taser,kim2010retro,hsu2006BFF,Webster2018criumr}.
If the recovery mechanism restores
the system to a sane state, the system continues to run with the same vulnerabilities and nothing stops
attackers
from reinfecting it.
Thus, the system could enter a loop of infections and recoveries.

Existing intrusion response systems, on the other hand,
apply responses~\cite{foo2005adepts}
to stop an intrusion or limit its impact on the system.
However, existing approaches
apply coarse-grained responses that affect the whole system and not just
the compromised services~\cite{foo2005adepts}
(\eg blocking port 80 for the whole system because a single compromised service
uses this port maliciously).
They also rely on a strong assumption of having complete knowledge of the vulnerabilities present
and used by the attacker~\cite{foo2005adepts,shameli2018dynamic} to select responses.

These limitations mean that they cannot respond to intrusions
without affecting the availability of the system or
of some services.
Whether it is due to business continuity, safety reasons, or the user experience,
the availability of services is an important aspect of a computing platform.
For example, while web sites, code repositories, or databases,
are not safety-critical, they can be important for a company or for the workflow of a user.
Therefore, the problem that we address is the following:
how to design an \gls{OS} so that its services can survive ongoing intrusions
while maintaining availability?

Our approach distinguishes itself from prior work on three fronts.
First, we \emph{combine} the restoration of files and processes
of a service with the ability to apply responses after
the restoration to withstand a reinfection.
Second, we apply \emph{per-service} responses that
affect the compromised services instead of the whole system
(\eg only one service views the file system as read-only).
Third, after recovering a compromised service,
the responses we apply can put the recovered service into a \emph{degraded mode},
because they remove some privileges normally needed by the service.

The degraded mode is on purpose.
When the intrusion is detected,
we do not have precise information about the vulnerabilities exploited
to patch them, or we do not have a patch available.
The degraded mode allows the system to survive the intrusion for two reasons.
First, after the recovery, the degraded mode either stops the attackers from reinfecting the service,
or it stops the attackers from achieving their goals.
Second, the degraded mode keeps as many functions of the service available as possible,
thus maintaining availability while waiting for a patch.

We maintain the availability by ensuring that core functions
of services are still operating, while non-essential functions might not be
working due to some responses.
For example, a web server could have "provide read access to the website" as core function,
and "log accesses" as non-essential.
Thus, if we remove the write access to the file system it would degrade the service's state (\ie it cannot log anymore),
but we would still maintain its core function.
We developed a cost-sensitive response selection
where administrators describe a policy consisting of
cost models for responses and malicious behaviors.
Our solution then selects
a response that maximize the effectiveness while minimizing its impact on the
service based on the policy.

This approach gives time for administrators to plan an update to fix the vulnerabilities
(\eg wait for a vendor to release a patch).
Finally, once they patched the system, we can remove the responses,
and the system can leave the degraded mode.

\noindent\textbf{Contributions.} Our main contributions are the following:
\begin{itemize}
    \item We propose a novel intrusion survivability approach to withstand ongoing intrusions
          and maintain the availability of core functions of services
          (section \ref{sec:approach} and \ref{sec:architecture}).
    \item We introduce a cost-sensitive response selection process to help select optimal responses
          (\autoref{sec:framework}).
    \item We develop a Linux-based prototype implementation by
          modifying the Linux kernel,
          systemd~\cite{systemd}, CRIU~\cite{CRIU}, Linux audit~\cite{auditd},
          and snapper~\cite{snapper} (\autoref{sec:implementation}).
    \item We evaluate our prototype by measuring the effectiveness of
          the responses applied, the ability to select appropriate responses,
          the availability cost of a checkpoint
          and a restore, the overhead of our solution, and
          the stability of the degraded services (\autoref{sec:evaluation}).
\end{itemize}

\noindent\textbf{Outline.} The rest of this document is structured as follows.
First, in~\autoref{sec:related_work}, we review
the state of the art on intrusion recovery and response systems.
In~\autoref{sec:scope}, we give an overview of our approach, and we define the scope of our work.
In~\autoref{sec:architecture}, we specify the requirements and architecture of our approach.
In~\autoref{sec:framework}, we describe how we select cost-sensitive responses and maintain core functions.
In~\autoref{sec:implementation}, we describe a prototype implementation which we
then evaluate in~\autoref{sec:evaluation}.
In~\autoref{sec:discussion}, we discuss some limitations of our work.
Finally, we conclude and give the next steps regarding our work in~\autoref{sec:conclusion}.

\section{Related Work}
\label{sec:related_work}
Our work is based on the concept of survivability~\cite{knight2003towards},
and specifically intrusion survivability,
since our approach focuses on withstanding ongoing intrusions.
We make a trade-off between the availability of the functionalities of a vulnerable service
and the security risk associated to maintaining them.
In this section, we review existing work on intrusion recovery and response systems.

\subsection{Intrusion Recovery Systems}
Intrusion recovery systems~\cite{ashvin2005taser,kim2010retro,hsu2006BFF,xiong2009shelf,Webster2018criumr} focus
on system integrity by recovering legitimate persistent data.
Except SHELF~\cite{xiong2009shelf} and CRIU-MR~\cite{Webster2018criumr},
the previous approaches do not preserve availability
since their restore procedure forces a system
shutdown, or they do not record the state of the processes.
Furthermore, except for CRIU-MR, they log all system events
to later replay legitimate operations~\cite{ashvin2005taser,kim2010retro,xiong2009shelf} or rollback
illegitimate ones~\cite{hsu2006BFF}, thus providing a fine-grained recovery.
Such an approach, however, generates gigabytes of logs per day inducing a high storage cost.

Most related to our work are SHELF~\cite{xiong2009shelf} and CRIU-MR~\cite{Webster2018criumr}.
SHELF recovers
the state of processes and identify infected files using a log of system events.
During recovery, it quarantines infected objects by freezing processes or forbidding access to files.
SHELF, however, removes this quarantined state as soon as it restores the system, allowing
an attacker to reinfect the system.
In comparison, our approach applies responses after a restoration
to prevent reinfection or the reinfected service to cause damages to the system.

CRIU-MR restores infected systems running within a Linux container.
When an \gls{IDS} notifies CRIU-MR that a container is infected,
it checkpoints the container's state (using CRIU~\cite{CRIU}), and
identifies malicious objects (\eg files) using
a set of rules.
Then, it restores the container while omitting
the restoration of the malicious objects.
CRIU-MR differs from other approaches, including ours, because it
uses a checkpoint immediately followed by a restore, only
to remove malicious objects.

The main limitation affecting prior work, including SHELF and CRIU-MR,
is that they neither prevent the attacker
from reinfecting the system nor
allow the system to withstand a reinfection since vulnerabilities are still present.

\subsection{Intrusion Response Systems}
Having discussed systems that recover from intrusions
and showed that it is not enough to withstand them,
we now discuss intrusion response
systems~\cite{balepin2003IDSAutomatedResponses,foo2005adepts,shameli2018dynamic}
that focus on applying responses to limit
the impact of an intrusion.

One area of focus of prior work is on how to model intrusion damages, or response costs,
to select responses.
Previous approaches either rely on directed graphs about system resources and cost
models~\cite{balepin2003IDSAutomatedResponses},
on attack graphs~\cite{foo2005adepts},
or attack defense trees~\cite{shameli2018dynamic}.
\citet{shameli2018dynamic} use \gls{MOO} methods to
select an optimal response based on such models.

A main limitation, and difference with our work, is that
these approaches
apply system-wide or coarse-grained responses
that affect every application in the \gls{OS}.
Our approach is more fine-grained, since we select and apply per-service responses that only
affect the compromised service,
and not the rest of the system.
Moreover, these approaches
cannot restore a service to a sane state.
Our approach, on the other hand, combines the ability
to restore and to apply cost-sensitive per-service responses.

Some of these approaches~\cite{foo2005adepts,shameli2018dynamic} also rely on the knowledge
of vulnerabilities present on the system and assume that the attacker can only exploit these
vulnerabilities.
Our approach does not rely on such prior knowledge, but relies on the knowledge that an intrusion
occurred, and the malicious behaviors exhibited by this intrusion.

\citet{huang2016talos} proposed a closely related approach that mitigates the impact of waiting
for patches when a vulnerability is discovered.
However, their system is not triggered by an \gls{IDS} but only by the discovery of
a vulnerability.
They instrument or patch vulnerable applications so that they do not execute their
vulnerable code, thus losing some functionality (similar to a degraded state).
They generate workarounds that minimize the cost of losing a functionality
by reusing error-handling code already present.
In our work, however, we do not assume any knowledge about the vulnerabilities,
and we do not patch or modify applications.
\section{Problem Scope}
\label{sec:scope}
\begin{figure}[ht]
    \centering
    \resizebox{\columnwidth}{!}{
        \tikzset{
    function/.style = {rounded corners, drop shadow, text centered, fill=white, draw},
    service/.style = {draw, rounded corners, minimum height=5em, minimum width=4.5em, thick},
    every point/.style = {circle, inner sep=0, line width=100em, minimum size=0.5em, opacity=1, draw, solid, fill=white},
    point/.style={insert path={node[every point, #1]{}}},
    point/.default={},
    processtree/.style={
        every child node/.style={anchor=west},
        grow via three points={one child at (0.5em,-0.7em) and two children at (0.3em,-0.7em) and (0.3em,-1.5em)},
        edge from parent path={(\tikzparentnode.south) |- (\tikzchildnode.west)}
    },
    process/.style={draw, minimum size=0.5em, scale=0.7},
    kernelobject/.style={draw, minimum size=0.4em, scale=0.5, regular polygon, regular polygon sides=3},
    kernel/.style = {minimum height=5em, rounded corners, thick, rectangle split, rectangle split parts=3, draw},
    subsystem/.style = {rounded corners, thick, rectangle split, rectangle split parts=3, draw},
    database/.style={
        cylinder,
        cylinder uses custom fill,
        cylinder body fill=white,
        cylinder end fill=white,
        shape border rotate=90,
        aspect=0.25,
        drop shadow={on layer=background},
        draw
    }
}
\def\mystrut{\vrule height 0em depth 0.4em width 0em}
\begin{tikzpicture}[-{Latex[length=0.6em]}, auto]
    % Processes
    \node[process] (processtree) {}
    child[-,processtree] { node[process] (processtree1) {}}
    child[-,processtree] { node[process] (processtree2) {}};
    \node[service, fit={(processtree) (processtree1) (processtree2)}] (service) {};

    % Service
    \node[below right] at (service.north west) {Service};

    % Control Points & Kernel
    \node[right=1.5em of service, draw, minimum width=0.9em, minimum height=5em, very thin] (interface) {};
    \node[kernel,right=1.5em of interface] (kernel) {Devices\mystrut\nodepart{two}Network\mystrut\nodepart{three}Filesystem\mystrut};

    % Platform
    \begin{scope}[on background layer]
        \node[draw, fit={(processtree) (processtree1) (processtree2) (service) (interface) (kernel)},inner sep=0.7em, xshift=0.1em, yshift=0.45em, fill=white, drop shadow] (OS) {};
        \node[below right] at (OS.north west) {Operating System};
    \end{scope}

    % Detection & Analysis
    \node[function, left=7em of OS.north west, yshift=-1.5em, align=center, preaction={fill=white}, pattern=north west lines] (detection) {Intrusion\\Detection};
    \node[fill=white, inner sep=1pt, align=center, yshift=0.55em] at (detection.center) {Intrusion};
    \node[fill=white, inner sep=1pt, align=center, yshift=-0.55em] at (detection.center) {Detection};

    % Reasoning
    \node[function, below left=1em of OS, align=left, xshift=-0.5em, text width=11em] (response)
    {{Recovery \& Response}\\[0.4em]\hrule height 0.1em
        \vspace{0.6em}
        1. Restore infected objects
        \\
        2. Maintain core functions
        \\
        3. Withstand reinfection
    };

    \node[database, above right=3em of response.north, align=center] (policies) {Policies};
    \node[fill=white, inner sep=1pt] at (policies.center) {Policies};

    % Arrows
    \draw[line width=0.1em] (detection) -- node[above] {Monitor} (detection.east -| OS.west);
    \draw[line width=0.1em] (detection) -- node[left,align=center] {Alert} (detection |- response.north);

    \draw[line width=0.1em] (response) -| node[above left, align=center] {Restore\\service} (service.south);
    \draw[line width=0.1em] (response) -| node[above left, yshift=-0.2em, align=center] {Apply\\respo-\\nses} (interface.south);
    \draw[line width=0.1em] (response) -| node[above left, align=center] {Restore\\files} (kernel.south);

    \draw[line width=0.1em] ([xshift=-0.5em]response.north west -| policies.south east) -- node[align=center, left] {Use} ([xshift=-0.5em]policies.south east);

    % Control points
    \draw[line width=0.1em, {Latex[length=0.6em]}-{Latex[length=0.6em]}] ([yshift=-0.5em]service.north east) -- ([yshift=-1.0em]interface.north) [point=ultra thick] -- ([yshift=-1.0em]kernel.north west);
    \draw[line width=0.1em, {Latex[length=0.6em]}-{Latex[length=0.6em]}] (service) -- (interface.center) [point=ultra thick] -- ([yshift=-2.5em]kernel.north west);
    \draw[line width=0.1em, {Latex[length=0.6em]}-] ([yshift=1.5em]service.south east) -- ([yshift=1.5em]interface.south) [point=ultra thick] -- ([yshift=1.5em]kernel.south west);
    \draw[line width=0.1em, -{Latex[length=0.6em]}] ([yshift=0.5em]service.south east) -- ([yshift=0.5em]interface.south) [point=ultra thick] -- ([yshift=0.7em]kernel.south west);

    \begin{customlegend}[legend cell align=left,
    legend style={at={(3.9,-2.8)}},
    legend entries={
        \footnotesize In scope,
        \footnotesize Out of scope
    }]
    \addlegendimage{legend image code/.code={\node[draw, minimum width=2em, minimum height=0.8em] {};}}
    \addlegendimage{legend image code/.code={\node[draw, minimum width=2em, minimum height=0.8em, pattern=north west lines] {};}}
    \end{customlegend}
\end{tikzpicture}
    }
    \Description{A figure describing succinctly and at a high-level our approach with four elements.
    First, the OS that executes services that use devices, the network, or the file system via the kernel.
    Second, the IDS (which is out of scope for our work) that monitors the services and that sends an alert when an intrusion is detected.
    Third, our main component that orchestrate the recovery and the responses, and it receives the
    alerts from the IDS.
    Finally, the policies that our main component uses to select the appropriate responses.}
    \caption{High-level overview of our intrusion survivability approach}
    \label{fig:approach}
\end{figure}
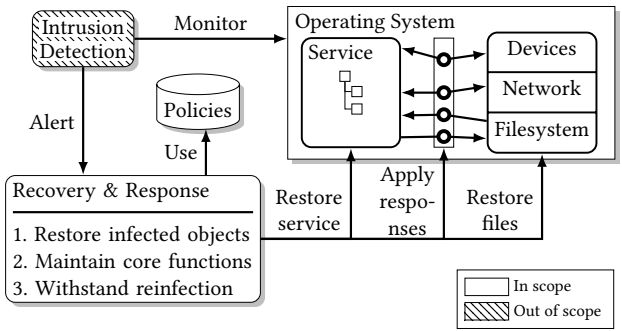

This section first gives an overview of our approach (illustrated in~\autoref{fig:approach}), then
it describes our threat model and some assumptions that narrow the attack scope.

\subsection{Approach Overview}
\label{sec:approach}
Since we focus our research on intrusion survivability, our work starts when
an \gls{IDS} detects an intrusion in a service.

When the \gls{IDS} detects an intrusion, we trigger a set of responses.
The procedure must meet the following goals:
restore infected objects (\eg files and processes),
maintain core functions, and
withstand a potential reinfection.
We achieve these goals using recoveries, responses, and policies.

\begin{description}
    \item[Recovery]
          Recovery actions restore the state of a service (\ie the state of its processes and metadata describing the service)
          and associated files to a previous safe state.
          To perform recovery actions, we create periodic snapshots of the filesystem and
          the services, during the normal operation of the \gls{OS}.
          We also log all the files modified by the monitored services.
          Hence, when restoring services, we only restore the files they modified.
          This limits the restoration time and it avoids the loss of
          known legitimate and non-infected data.%
          \footnote{Other files that the service depends on can be modified
           by another service, we handle such a case with dependencies information
           between services.}
          To perform recovery actions, we do not require for the system
          to be rebooted, and we limit the availability impact on the service.

    \item[Response]
          A response action removes privileges,
          isolates components of the system from the service,
          or reduces resource quotas (\eg CPU or RAM) of one service.
          Hence, it does not affect any other component of the system (including other services).%
          \footnote{However, if components depend on a degraded service, they can be affected indirectly.}
          Its goal is to either prevent an attacker to reinfect the service or
          to withstand a reinfection
          by stopping attackers from achieving their goals
          (\eg data theft) after the recovery.
          Such a response may put the service into a degraded mode,
          because some functions might not have the
          required privileges anymore (or limited access to some resources).

    \item[Policies]
          We apply appropriate responses that do not disable core functions
          (\eg the ability to listen on port 80 for a web server).
          To refine the notion of core functions, we rely on policies.
          Their goal is to provide a trade-off between
          the availability of a function (that requires specific privileges)
          in a service and the intrusion risk.
          We designed a process
          to select cost-sensitive responses (see~\autoref{sec:framework})
          based on such policies.
          Administrators, developers, or maintainers provide the policies
          by specifying the cost of losing
          specific privileges (\ie if we apply a response) and the cost of
          a malicious behavior (exhibited by an intrusion).
\end{description}

\subsection{Threat Model and Assumptions}
We make assumptions regarding the platform's firmware (\eg BIOS or UEFI-compliant firmware)
and the \gls{OS} kernel where we execute the services.
If attackers compromise such components at boot time or runtime,
they could compromise the \gls{OS} including our mechanisms.
Hence, we assume their integrity.
Such assumptions are reasonable in recent firmware using
a hardware-protected root of trust~\cite{intelBootGuard,hpSureStart} at boot time
and protection of firmware runtime
services~\cite{yao2015whiteMemoryProtection,yao2015whiteSMMMonitor,chevalier2017coproc}.
For the \gls{OS} kernel,
one can use UEFI Secure Boot~\cite{UEFISpec} at boot time, and
rely on \eg security invariants~\cite{song2016enforcing} or
a hardware-based integrity monitor~\cite{azab2014hypervision} at runtime.
The main threat that we address is the compromise of services inside an \gls{OS}.

We make no assumptions regarding the privileges that were initially granted to the services.
Some of them can restrict their privileges to the minimum.
On the contrary, other services can be less effective in adopting the principle of least privilege.
The specificity of our approach is that we deliberately remove privileges that could not
have been removed initially, since the service needs them for a function it provides.
Finally, we assume that the attacker cannot compromise the mechanisms we use to
checkpoint, restore, and apply responses
(\autoref{sec:architecture} details how we protect such mechanisms).

We model an attacker with the following capabilities:
\begin{itemize}
    \item Can find and exploit a vulnerability in a service,
    \item Can execute arbitrary code in the same context as the compromised service,
    \item Can perform some malicious behaviors even if the service had initially the minimum amount of privileges to accomplish its functions,
    \item Can compromise a privileged service or elevate the privileges of a compromised service to superuser,
    \item Cannot exploit software-triggered hardware vulnerabilities (\eg side-channel attacks~\cite{kim2014flipping,seaborn2015exploiting,Kocher2018spectre,Lipp2018meltdown}),
    \item Do not have physical access to the platform.
\end{itemize}

\section{Architecture and Requirements}
\label{sec:architecture}
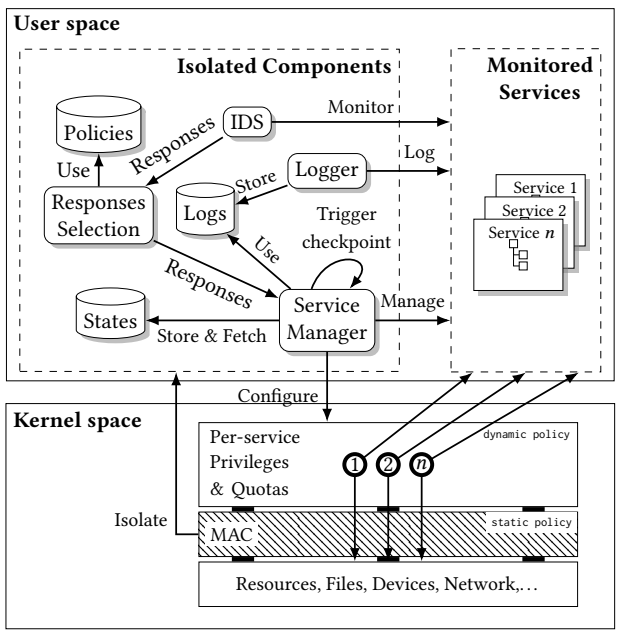
\begin{figure}[ht]
    \centering
    \resizebox{\columnwidth}{!}{
        \tikzset{
    interface/.style = {draw, minimum width=17em, minimum height=2em, align=center},
    function/.style = {rounded corners, drop shadow, text centered, fill=white, draw},
    block/.style = {ellipse, draw, text centered, fill=white},
    service/.style = {draw, fill=white, drop shadow, minimum height=2em, inner sep=0.9em, minimum width=4em},
    every point/.style = {circle, line width=100em, minimum size=0.5em, opacity=1, draw, solid, fill=white},
    point/.style={insert path={node[every point, #1]{}}},
    point/.default={},
    processtree/.style={
        every child node/.style={anchor=west},
        grow via three points={one child at (0.5em,-0.7em) and two children at (0.5em,-1.0em) and (0.5em,-2em)},
        edge from parent path={(\tikzparentnode.south) |- (\tikzchildnode.west)}
    },
    process/.style={draw, minimum size=0.2em, scale=0.2},
    database/.style={
        cylinder,
        cylinder uses custom fill,
        cylinder body fill=white,
        cylinder end fill=white,
        shape border rotate=90,
        aspect=0.25,
        drop shadow={on layer=background},
        draw
    }
}
\def\mystrut{\vrule height 0em depth 0.4em width 0em}
\begin{tikzpicture}[-{Latex[length=0.6em]}, auto]
    \begin{scope}[local bounding box=Userspace]
        \begin{scope}[local bounding box=IsolatedComponents]
            \node[function, align=center] (SM) {Service\\Manager};
            \node[database, left=6em of SM] (StatesStorage) {States};
            
            \node[function, above=4.5em of SM] (Logger) {Logger};
            \node[database, above left=3.5em of SM] (LogsStorage) {Logs};
            
            \node[function, left=1em of LogsStorage,align=center] (RS) {Responses\\Selection};
            \node[function, above left=1em of Logger] (IDS) {IDS};
            
            \node[database, above=1.5em of RS] (ModelsStorage) {Policies};
        \end{scope}
    
        \node[draw, dashed, fit={(IsolatedComponents)}, inner sep=1em, inner xsep=1.1em, inner ysep=1.5em, yshift=0.6em, minimum width=6em] (IsolatedComponentsBox) {};
        \node[below left] at (IsolatedComponentsBox.north east) {\textbf{Isolated Components}};
    
        \begin{scope}[shift={($(IsolatedComponentsBox.east) + (2,0)$)}, local bounding box=MonitoredServices, on background layer]
            \node[service] (S1) {
                \begin{tikzpicture}[-, scale=0.5, processtree]
                \node[process] (processtree) {}
                child { node[process] (processtree1) {}}
                child { node[process] (processtree2) {}};
                \end{tikzpicture}
            };
            \node[service, xshift=-0.5em, yshift=-1em] (S2) {
                \begin{tikzpicture}[-, scale=0.5, processtree]
                \node[process] (processtree) {}
                child { node[process] (processtree1) {}}
                child { node[process] (processtree2) {}};
                \end{tikzpicture}
            };
            \node[service, xshift=-1em, yshift=-2em] (SN) {
                \begin{tikzpicture}[-, scale=0.5, processtree]
                \node[process] (processtree) {}
                child { node[process] (processtree1) {}}
                child { node[process] (processtree2) {}};
                \end{tikzpicture}
            };
    
            \node[below left] at (SN.north east) {\footnotesize Service $n$};
            \node[below left] at (S2.north east) {\footnotesize Service $2$};
            \node[below left] at (S1.north east) {\footnotesize Service $1$};
        \end{scope}
        \node[draw, dashed, fit={(MonitoredServices) (IsolatedComponentsBox.north east) (IsolatedComponentsBox.south east)}, inner sep=0pt, inner xsep=-0.8em, xshift=1.5em] (MonitoredServicesBox) {};
        \node[below left, align=center] at (MonitoredServicesBox.north east) {\textbf{Monitored}\\\textbf{Services}};

    \end{scope}
    \node[draw, fit={(Userspace)}, inner sep=0.6em, inner ysep=1.1em, yshift=0.7em] (UserspaceBox) {};
    \node[below right] at (UserspaceBox.north west) {\textbf{User space}};

    \begin{scope}[shift={($(UserspaceBox.south) + (0,-1.5)$)}, local bounding box=Kernel]
        \node[interface, align=left, text width=16em, xshift=3.5em, yshift=1em] (CP) {\small Per-service\\\small Privileges\\\small\& Quotas};
        \node[below left] at (CP.north east) {\tiny \texttt{dynamic policy}};
        \node[interface, pattern=north west lines, align=left, text width=16em, below=0.2em of CP] (MAC) {};
        \node[right, fill=white, xshift=0.3em, inner sep=2pt] at (MAC.west) {\small MAC};
        \node[below left, fill=white, inner sep=2pt, yshift=-0.03em, xshift=-0.03em] at (MAC.north east) {\tiny \texttt{static policy}};
        \node[interface, below=0.2em of MAC] (Resources) {\small Resources, Files, Devices, Network,\ldots};
    \end{scope}
    \node[draw, fit={(Kernel) (UserspaceBox.south east) (UserspaceBox.south west)}, inner sep=0pt, yshift=-1em] (KernelBox) {};
    \node[below right] at (KernelBox.north west) {\textbf{Kernel space}};

    \draw[line width=0.08em] ([xshift=-0.7em]SM.north) to [out=70, in=50, looseness=3] node[above, align=center] {\small Trigger\\\small checkpoint} ([xshift=1em]SM.north);

    \draw[line width=0.08em] (RS) -- node[left] {Use} (ModelsStorage);
    \draw[line width=0.08em] (IDS) -- node[above,sloped] {Responses} (RS);
    \draw[line width=0.08em] (RS) -- node[below,sloped] {Responses} (SM);
    \draw[line width=0.08em] (IDS) -- node[above] {\small Monitor} (IDS.east -| MonitoredServicesBox.west);
    \draw[line width=0.08em] (Logger) -- node[above, sloped] {\small Store} (LogsStorage);
    \draw[line width=0.08em] (Logger) -- node[above, xshift=0.5em] {\small Log} (Logger.east -| MonitoredServicesBox.west);
    \draw[line width=0.08em] (SM) -- node[below] {\small Store \& Fetch} (StatesStorage);
    \draw[line width=0.08em] (SM) -- node[above] {\small Manage} (SM.east -| MonitoredServicesBox.west);
    \draw[line width=0.08em] (SM) -- node[above, sloped] {\small Use} (LogsStorage);
    
    \draw[line width=0.08em] (MAC) -| node[above left] {\small Isolate} ([xshift=-1.5em]IsolatedComponentsBox.south);
    \draw[line width=0.08em] (SM) -- node[left, yshift=-0.49em] {\small Configure} (CP.north -| SM.south);
    
    % Control points
    \draw[line width=0.08em, {Latex[length=0.6em]}-{Latex[length=0.6em]}] ([xshift=1em]MonitoredServicesBox.south west) -- ([xshift=-1.5em]CP.center) node[point=ultra thick] {1} -- ([xshift=-1.5em]Resources.north);
    \draw[line width=0.08em, {Latex[length=0.6em]}-{Latex[length=0.6em]}] (MonitoredServicesBox.south) -- (CP.center) node[point=ultra thick] {2} -- (Resources);
    \draw[line width=0.08em, {Latex[length=0.6em]}-{Latex[length=0.6em]}] ([xshift=-1em]MonitoredServicesBox.south east) -- ([xshift=1.5em]CP.center) node[point=ultra thick] {$n$} -- ([xshift=1.5em]Resources.north);
    
    % Links between kernel layers
    \draw[-, line width=1em] (CP) -- (MAC);
    \draw[-, line width=1em] ([xshift=2em]CP.south west) -- ([xshift=2em]MAC.north west);
    \draw[-, line width=1em] ([xshift=-2em]CP.south east) -- ([xshift=-2em]MAC.north east);

    \draw[-, line width=1em] (MAC) -- (Resources);
    \draw[-, line width=1em] ([xshift=2em]MAC.south west) -- ([xshift=2em]Resources.north west);
    \draw[-, line width=1em] ([xshift=-2em]MAC.south east) -- ([xshift=-2em]Resources.north east);
\end{tikzpicture}
    }
    \Description{A figure that provides an overview of our architecture.
    It is divided it two parts: user space and kernel space.
    In user space, we have our isolated components and the monitored services.
    In kernel space, we have the various elements that allows the kernel to enforce
    a dynamic policy on the services configured by the service manager from user space.
    It also enforces a static policy to isolate our user space components.
    Our isolated components include the IDS, the logging facility, the service manager,
    the response selection, the policies, the logs, and the states.}
    \caption{Overview of the architecture}
    \label{fig:architecture_overview}
\end{figure}

Our approach relies on four components.
In this section, we first give an overview of how each component works and interacts with
the others, as illustrated in~\autoref{fig:architecture_overview}.
Then, we detail requirements about our architecture.

\subsection{Overview}

During the normal operation of the \gls{OS}, the \emph{service manager} creates periodic
checkpoints of the services and snapshots of the filesystem.
In addition, a \emph{logging facility} logs the path of all the files modified
by the monitored services since their last checkpoint.
The logs are later used to filter the files that need to be restored.

The \emph{\gls{IDS}} notifies the \emph{responses selection} component when it
detects an intrusion
and specifies information about possible responses to withstand it.
The selected responses are then given to the service manager.
The service manager
restores the infected service to the last known safe state
including all the files modified by the infected service.
Then, it configures kernel-enforced per-service privilege restrictions and quotas
based on the selected responses.
To select the last known safe state, we rely on the \gls{IDS} to identify the first
alert related to the intrusion.
Then, we consider that the first state prior to this alert is safe.

\subsection{Isolation of the Components}
For our approach to be able to withstand an attacker trying to impede the detection
and recovery procedures, the integrity and availability of each component is crucial.
Different solutions (\eg a hardware isolated execution environment or a hosted hypervisor) could
be used.
In our case, we rely on a kernel-based \gls{MAC} mechanism, such as SELinux~\cite{SELinux},
to isolate
the components we used in our approach.
Such a mechanism is available in commodity \glspl{OS},
can express our isolation requirements,
and does not modify the applications.
We now give guidelines on how to build a \gls{MAC} policy to
protect our components.

First, the \gls{MAC} policy must ensure that none of our components can be killed.%
\footnote{One can also use a watchdog to ensure that the
    components are alive.}
Otherwise, \eg if the responses selection component is not alive,
no responses will be applied.

Second, the \gls{MAC} policy must ensure that only our components have access
to their isolated storage (\eg to store the logs or checkpoints).
Otherwise, attackers might \eg erase an entry to avoid
restoring a compromised file.

Third, the \gls{MAC} policy must restrict the communication between the different components,
and it must only allow a specific program to advertise itself as one of the components.
Otherwise, an attacker might impersonate a component or stop the communication between
two components.
In our case, we assume a \gls{RPC} or
an \gls{IPC} mechanism that can implement \gls{MAC} policies
(\eg D-Bus~\cite{dbus} is SELinux-aware~\cite{SELinuxAware}).

\subsection{Intrusion Detection System}
\label{sec:architecture:ids}
Our approach requires an \gls{IDS} to detect an intrusion
in a monitored service.
We do not require a specific type of \gls{IDS}.
It can be external to the system or not.
It can be misuse-based or anomaly-based.
We only have two requirements.

First, the \gls{IDS} should be able to pinpoint the intrusion to a specific service
to apply per-service responses.
For example, if the \gls{IDS} analyzes event logs to detect intrusions,
they should include the service that triggered the event.

Second, the \gls{IDS} should have information about the intrusion.
It should map the intrusion to a set of malicious behaviors
(\eg the malware capabilities~\cite{maec_hierarchy} from \gls{MAEC}~\cite{maec}),
and it should provide a set of responses that
can stop or withstand them.
Both types of information
can either be part of the alert from the \gls{IDS} or
be generated from threat intelligence based on the alert.
Generic responses can also be inferred due to the type of intrusion
if the \gls{IDS} lacks precise information about the intrusion.
For example, a generic response for ransomware consists in
setting the filesystem hierarchy as read-only.
Information about the alert, the responses, or
malicious behaviors,
can be shared using standards such as \gls{STIX}~\cite{stix} and \gls{MAEC}~\cite{maec,EMA}.

\subsection{Service Manager}
Commodity \glspl{OS} rely on a user space service manager
(\eg the Service Control Manager~\cite{windowsInternalsPart2} for Windows,
or systemd~\cite{systemd} for Linux distributions)
to launch and manage services.
In our architecture, we rely on it, since
it provides the appropriate level of abstraction
to manage services and it has the notion of dependencies
between services.
Using such information,
we can restore services in a coherent state.
If a service depends on other services (\eg if one service writes to a file and another one reads it),
we checkpoint and restore them together.

We extend the service manager to checkpoint and restore the state
of services.
Furthermore,
we modify the service manager so that it applies responses
before it starts a recovered service.
Since such responses are per-service,
the service manager must have access
to \gls{OS} features to configure per-service privileges
and resource quotas.

The service manager must be able to kill a service (\ie all alive processes
created by the service) if it is compromised and needs to be restored.
Therefore, we bound processes to the service that created them,
and they must not be able to break the bound.
For example, we can use cgroups~\cite{cgroupsv2} in Linux or job
objects~\cite{MSDNJobObjects} in Windows.

Finally, the \gls{MAC} policy must ensure that only the service manager manages
the collections of processes
(\eg \verb|/sys/fs/cgroup| in Linux).
Otherwise, if an attacker breaks the bound of a compromised service, it would be
difficult to kill the escaped processes.
Likewise, the \gls{MAC} policy must protect configuration files
used by the service manager.
\section{Cost-Sensitive Response Selection}
\label{sec:framework}
For a given intrusion, multiple responses might be appropriate, and each one incurs an availability cost.
We devised a framework to help select the cost-sensitive responses that minimize such a cost and
maintain the core functions of the service.

We use a qualitative approach using linguistic constants (\eg low or high) instead of a
quantitative one (\eg monetary values).
Quantitative approaches require an accurate value of assets, and historical data of previous
intrusions to be effective, which we assume missing.
Qualitative approaches, while prone to biases and inaccuracies, do not require
such data, and are easier to understand~\cite{wheeler2011securityChapter2}.
In addition, we would like to limit the input from the user
so that it improves the framework usability and its likelihood to be adopted in production.

In the rest of this section, we first describe the models that our framework relies on.
Then, we detail how our framework selects cost-sensitive responses using such models.

\subsection{Malicious Behaviors and Responses}
\label{sec:policies:malicious_behaviors_and_responses}
Intrusions may exhibit multiple malicious behaviors that need to be stopped
or mitigated differently.
Here we work at the level of a malicious behavior
and we select a response for each malicious behavior.

Our models rely on a hierarchy of malicious behaviors where the first levels describe high-level
behaviors (\eg compromise data availability), while lower levels describe more precise behaviors
(\eg encrypt files).
The malware capabilities hierarchy~\cite{maec_hierarchy} from the project \gls{MAEC}~\cite{maec}
of MITRE is a suitable candidate for such a hierarchy.%
\footnote{Another project that can help is the MITRE ATT\&CK knowledge base~\cite{attck},
but it does not provide a hierarchy.}
We model this hierarchy as a partially ordered set ($\mathbf{M}, \prec_{\mathbf{M}}$) with
$\prec_{\mathbf{M}}$ a binary relation over the set of malicious behaviors $\mathbf{M}$.
The relation
$m \prec_{\mathbf{M}} m^\prime$ means that $m$ is a more precise behavior than $m^\prime$.
Let $\mathbf{I}$ be the space of intrusions reported by the \gls{IDS}.
We assume that for each intrusions $i \in \mathbf{I}$, we can
map the set of malicious behaviors  $M^i \subseteq \mathbf{M}$ exhibited by $i$.
By construct, we have the following property:
if $m \prec_{\mathbf{M}} m^\prime$
then $m \in M^i \implies m^\prime \in M^i$.

We also rely on a hierarchy of responses where the first levels describe
coarse-grained responses (\eg block the network), while lower levels describe
more fine-grained responses (\eg block port 80).
We define the hierarchy as a partially ordered set ($\mathbf{R}, \prec_{\mathbf{R}}$) with
$\prec_{\mathbf{R}}$ a binary relation over the set of responses $\mathbf{R}$
($r \prec_{\mathbf{R}} r^\prime$ means that $r$ is a more fine-grained response than $r^\prime$).
Let $R^m \subseteq \mathbf{R}$ be the set of responses that can stop
a malicious behavior $m$.
By construct, we have the following property:
if $r \prec_{\mathbf{R}} r^\prime$
then $r \in R^m \implies r^\prime \in R^m$.
Such responses are based on the OS-features available to restrict
privileges and quotas on the system.
We provide an example of this response hierarchy
in~\autoref{fig:response_hierarchy_example} of \aref{appendix:models}.

\subsection{Cost Models}
Let the space of services be denoted $\mathbf{S}$
and let the space of qualitative linguistic constants be a totally ordered set, denoted $\mathbf{Q}$
composed as follows:
$\text{none} < \text{very low} < \text{low} < \text{moderate} < \text{high} < \text{very high} < \text{critical}$.
We extend each service configuration file with the notion of response cost
(in terms of quality of service lost) and malicious behavior cost
that an administrator needs to set.

A \textbf{response cost} $c_r \in \mathbf{C_{r}} \subseteq \mathbf{Q}$ is the qualitative impact of applying a response
$r \in \mathbf{R}$ on a service to
stop a malicious behavior.
We define $rcost : \mathbf{S} \times \mathbf{R} \rightarrow \mathbf{C_{r}}$,
the function that takes a service, a response, and returns the associated response cost.

Response costs allow an administrator or developer of a service to
specify how a response, if applied, would impact the
overall quality of service.
The impact can be assessed based on the number of functions that would be unavailable
and their importance for the service.
More importantly, with the value critical, we consider that a response
would disable a core function of a service and thus should never be applied.

For example, the policy of a web server could express that the ability to listen on ports
80 and 443 is critical for its core functions.
However, if the web server would lose write access to the filesystem,
the cost would be high and not critical,
since it can still provide access to websites for many use cases.

A \textbf{malicious behavior cost} $c_{mb} \in \mathbf{C_{mb}} \subseteq \mathbf{Q}$ is the qualitative impact of a malicious
behavior $m \in \mathbf{M}$.
We define $mbcost \colon \mathbf{S} \times \mathbf{M} \to \mathbf{C_{mb}}$,
the function that takes a service, a malicious behavior, and returns the associated cost.

We require for each service that a malicious behavior cost is set
for the first level of the malicious behaviors hierarchy (\eg there are only 20 elements on the first level
of the hierarchy from \gls{MAEC}).
We do not require it for other levels, but if more costs are set, then the response
selection will be more accurate.
The $mbcost$ function associates a cost for each malicious behavior $m$.
The cost, however, could be undefined.
In such a case, we take the cost of $m^\prime$ such that
$mbcost(s, m^\prime)$ is defined, $m \prec_{\mathbf{M}} m^\prime$,
and $\nexists\,m^{\prime\prime}$ such that $m < m^{\prime\prime} < m^\prime$
with $mbcost(s, m^{\prime\prime})$ defined.

Following the same example, the policy could express that intrusions
that compromise data availability (\eg ransomware) have a high impact for the web server,
since it would not provide access to the websites anymore.
While on the other hand, it could express that an intrusion that only consumes system resources
(\eg a cryptocurrency mining malware) has a moderate cost.

Both costs need to be configured depending on the \emph{context} of the service.
For example, a web server that provides static content does not have the same context,
hence the same costs than one that handles transactions.

\subsection{Response Performance}
While responses have varying costs on the quality of service,
they also differ in performance against a malicious behavior.
Hence, in our framework, we consider the performance as a criterion to select
a response, among others.%
\footnote{The most effective response would be to stop the service.
While our model allows it, in this paper we only mention responses that aim at
maintaining the availability.}

The space of qualitative response performances is denoted $\mathbf{P_{r}} \subseteq \mathbf{Q}$.
We define $rperf \colon \mathbf{R} \times \mathbf{M} \to \mathbf{P_{r}}$,
that takes a response, a malicious behavior, and returns the associated performance.

In contrast to the cost models previously defined
that are specific to a system and its context
(and need to be set, \eg by an administrator of the system),
such a value only depends on the malicious behavior and is provided by security
experts that analyzed similar intrusions and proposed
responses with their respective performance.
Such information comes from threat intelligence sources that are shared, for example,
using \gls{STIX}.
For example, \gls{STIX} has a property called "efficacy" in its "course-of-action" object that
represent responses.

\subsection{Risk Matrix}
We rely on the definition of a risk matrix that
satisfies the axioms proposed
(\ie weak consistency, betweenness, and consistent coloring)
to provide consistent risk assessments~\cite{cox2008matrices}.
The risk matrix needs to be defined ahead of time by the administrator
depending on the risk attitude of the organization:
whether the organization is risk averse, risk neutral, or risk seeking.
The $5 \times 5$  risk matrix shown in~\autoref{tbl:risk_matrix} of~\aref{appendix:risk_matrix}
is one instantiation of such a matrix.

The risk matrix outputs a qualitative malicious behavior risk
$k \in \mathbf{K} \subseteq \mathbf{Q}$.
The risk matrix depends on a malicious behavior cost (impact),
and on the confidence level $i_{cf} \in \mathbf{I_{cf}} \subseteq \mathbf{Q}$
that the \gls{IDS} has on the intrusion (likelihood).

We define $risk \colon \mathbf{C_{mb}} \times \mathbf{I_{cf}} \to \mathbf{K}$,
the function representing the risk matrix that takes a malicious behavior cost,
an intrusion confidence,
and returns the associated risk.

\subsection{Policy Definition and Inputs}
Having discussed the various models we rely on,
we can define the policy as a tuple of four functions
$\langle rcost, rperf, mbcost, risk \rangle$.
The $risk$ function is defined at the organization level,
$mbcost$ and $rcost$ are defined for each
service depending on its context, and $rperf$ is constant and
can be applied for any system.
Hence, the most time-consuming parameters to set are $mbcost$ and $rcost$.

The function $mbcost$ can be defined by someone that understands the impact
of malicious behaviors based on the service's context (\eg an administrator).
$rcost$ can be defined by
an expert, a developer of the service, or a maintainer of the \gls{OS} where the service is used,
since they understand the impact of removing certain privileges to the service.
For example, some Linux distributions provide the security policies
(\eg SELinux or AppArmor) of their services and applications.
Much like SELinux policies, $rcost$ could be provided this way,
since the maintainers would need to test that the response do not render
a service unusable (\ie by disabling a core functionality).

\subsection{Optimal Response Selection}
We now discuss how we use our policy to select cost-sensitive responses.
Our goal is to maximize the performance of the response
while minimizing the cost to the service.
We rely on known \gls{MOO} methods~\cite{marler2004survey}
to select the most cost-effective response,
as does other work on response selection~\cite{shameli2018dynamic,motzek2017selection}.

For conciseness, since we are selecting a response for a malicious behavior
$m \in \mathbf{M}$
and a service $s \in \mathbf{S}$,
we now denote $rperf(r, m)$ as $p_r$,
$rcost(s, r)$ as $c_r$,
and $mbcost(s, m)$ as $c_{mb}$.

\subsubsection{Overview}
When the \gls{IDS} triggers an alert, it provides the confidence $i_{cf} \in \mathbf{I_{cf}}$
of the intrusion $i \in \mathbf{I}$
and the set of malicious behaviors $M^i \subseteq \mathbf{M}$.
Before selecting an optimal response, we filter out
any response that have a critical response cost from $R^m$ (the space of responses that can stop
a malicious behavior $m$).
Otherwise, such responses would impact a core function of the service.
We denote $\hat{R}^m \subseteq R^m$ the resulting set:
\begin{equation*}
\hat{R}^m = \{\, r \in R^m \mid c_r < \text{critical}\,\}
\end{equation*}

For each malicious behavior $m \in M^i$, we compute the Pareto-optimal set from $\hat{R}^m$,
where we select an optimal response from.
We now describe these last steps.

\subsubsection{Pareto-Optimal Set}
In contrast to a \gls{SOO} problem, a \gls{MOO} problem does not
generally have a single global solution.
For instance, in our case we might not have a response that provides both the maximum performance
and the minimum cost, because they are conflicting,
but rather a set of solutions that are defined as optimum.
A common concept to describe such solutions is Pareto optimality.

A solution is Pareto-optimal (non-dominated) if it is not possible to find other solutions
that improve one objective
without weakening another one.
The set of all Pareto-optimal solutions is called a Pareto-optimal set (or Pareto front).
More formally, in our context, we say that a response is Pareto-optimal if it is non-dominated.
A response $r \in R^m$ dominates a response $r' \in R^m$, denoted $r \succ r'$, if the following is satisfied:
\begin{equation*}
    [p_r > p_{r'} \wedge c_r \leq c_{r'}] \vee [p_r \geq p_{r'} \wedge c_r < c_{r'}]
\end{equation*}

\Gls{MOO} methods rely on preferences to choose solutions among the Pareto-optimal set
(\eg should we put the priority on the performance of the response or on reducing the cost?)~\cite{marler2004survey}.
They rely on a scalarization that converts a \gls{MOO} problem into a \gls{SOO} problem.
One common scalarization approach is the weighted sum method that assigns a weight to each objective and
compute the sum of the product of their respective objective.
However, this method is not guaranteed to always give solutions in the
Pareto-optimal set~\cite{marler2004survey}.

\citet{shameli2018dynamic} decided to apply the weighted sum method on the Pareto-optimal set instead
of on the whole solution space to guarantee to have a solution in the Pareto-optimal set.
We apply the same reasoning, so we reduce our set to all non-dominated responses.
We denote the resulting Pareto-optimal set $\mathcal{O}$:
\begin{equation*} \label{eq:pareto_optimal_set}
    \mathcal{O} = \{\, r_i \in \hat{R}^m \mid \nexists r_j \in \hat{R}^m, r_j \succ r_i\,\}
\end{equation*}

\subsubsection{Response Selection}
Before selecting a response from the Pareto-optimal set using use the weighted sum method,
we need to set weights,
and to convert the linguistic constants, we use to define the costs, into numerical values.

We rely on a function $l$ that maps the linguistic constants to a numerical value%
\footnote{An alternative would be to use fuzzy logic to reflect the uncertainty regarding
    the risk assessment from experts when using linguistic constants~\cite{deng2011risk}.}
between $0$ and $1$.
In our case, we convert the constants critical, very high, high, moderate, low, very low,
and none, to respectively the value 1, 0.9, 0.8, 0.5, 0.3, 0.1, and 0.

For the weights,
we use the risk $k = risk(c_{mb}, i_{cf})$
as a weight for the performance of the response $w_p = l(k)$
which also gives us the weight for the cost of a response $w_c = 1 - w_p$.
It means that we prioritize the performance if the risk is high, while we
prioritize the cost if the risk is low.

We obtain the final optimal response by applying the weighted sum method:
\begin{equation*} \label{eq:best}
    \argmax_{r \in \mathcal{O}} w_p l(p_r) + w_c (1 - l(c_r))
\end{equation*}

\section{Implementation}
\label{sec:implementation}
We implemented a Linux-based prototype by modifying several existing projects.
While our implementation relies on Linux features such as
namespaces~\cite{kerrisk2013namespaces},
seccomp~\cite{corbet2009seccomp},
or cgroups~\cite{cgroupsv2},
our approach does not depend on \gls{OS}-specific paradigms.
For example, on Windows, one could use Integrity Mechanism~\cite{MSDNWIM},
Restricted Tokens~\cite{MSDNRestrictedTokens}, and Job Objects~\cite{MSDNJobObjects}.
In the rest of this section, we describe the projects we modified,
why we rely on them, and the different modifications we made to implement our prototype.
You can see in~\autoref{tbl:code_added}
the different projects we modified where we added in total
nearly 3600 lines of C code.

\begin{table}[ht]
    \centering
    \caption{Projects modified for our implementation}
    \label{tbl:code_added}
    \begin{tabular}{rllr}
        \toprule
        \multicolumn{2}{l}{Project} & From version & Code added                       \\
        \midrule
        \multicolumn{2}{l}{CRIU}    & 3.9          & 383 lines of C                   \\
        \multicolumn{2}{l}{systemd} & 239          & 2639 lines of C                  \\
        \multicolumn{2}{l}{audit}   &              &                                  \\
                     & user space   & 2.8.3        & 79 lines of C                    \\
                     & Linux kernel & 4.17.5       & 460 lines of C                   \\
        \midrule
        \multicolumn{2}{l}{Total}   &              & 3561 lines of C                  \\
        \bottomrule
    \end{tabular}
\end{table}

At the time of writing, the most common service manager on Linux-based systems is systemd~\cite{systemd}.
We modified it to checkpoint and to restore services
using CRIU~\cite{CRIU} and snapper~\cite{snapper},
and to apply responses at the end of the restoration.
\subsection{Checkpoint and Restore}
CRIU is a checkpoint and restore project implemented in user space for Linux.
It can checkpoint the state of an application by fetching information
about it from different kernel APIs, and then store this information inside an image.
CRIU reuses this image and other kernel APIs to restore the application.
We chose CRIU because it allows us to perform transparent checkpointing and restoring
(\ie without modification or recompilation) of the services.

Snapper provides an abstraction for snapshotting filesystems
and handles multiple Linux filesystems (\eg BTRFS~\cite{rodeh2013btrfs}).
It can create a comparison between a snapshot and another one (or the current state of the filesystem).
In our implementation, we chose BTRFS due its \gls{COW} snapshot and comparison features,
allowing a fast snapshotting and comparison process.

When checkpointing a service, we first freeze its cgroup
(\ie we remove the processes from the scheduling queue)
to avoid inconsistencies.
Thus, it cannot interact with other processes nor with the filesystem.
Second, we take a snapshot of the filesystem and a snapshot of the metadata of the service kept
by systemd (\eg status information).
Third, we checkpoint the processes of the service using CRIU.
Finally, we unfreeze the service.

When restoring a service, we first kill all the processes belonging to its cgroup.
Second, we restore the metadata of the service
and ask snapper to create a read-only snapshot of the current state of the filesystem.
Then, we ask snapper to perform
a comparison between this snapshot and the snapshot taken during the checkpointing of
the service.
It gives us information about which files were modified and how.
Since we want to only recover the files modified by the monitored service,
we filter the result based on our log of files modified
by this specific service
(see~\autoref{sec:implementation:monitoring} for more details)
and restore the final list of files.
Finally, we restore the processes using CRIU.
Before unfreezing the restored service,
CRIU calls back our function that applies the responses.
We apply the responses at the end to avoid interfering with CRIU that requires certain
privileges to restore processes.

\subsection{Responses}
\label{sec:implementation:mitigations}
Our implementation relies on Linux features such as
namespaces,
seccomp,
or cgroups, to apply responses.
Here is a non-exhaustive list of responses
that we support:
filesystem constraints
(\eg put all or any part of the filesystem read-only),
system call filters
(\eg blacklisting a list or a category of system calls),
network socket filters (\eg deny access to a specific IP address), or
resource constraints (\eg CPU quotas or limit memory consumption).

We modified systemd to apply most of these responses just before unfreezing the restored service,
except for system call filters.
Seccomp only allows processes to set up their own filters and
prevent them to modify the filters of other processes.
Therefore, we modified systemd so that when CRIU restores a process,
it injects and executes
code inside the address space of the restored process to set up our filters.

\subsection{Monitoring Modified Files}
\label{sec:implementation:monitoring}
The Linux auditing system~\cite{auditd,auditgithub} is a standard way
to trigger events from the kernel to user space based on a set of rules.
Linux audit can trigger events when a process performs write accesses on the filesystem.
However, it cannot filter these events for a set of processes
corresponding to a given service (\ie a cgroup).
Hence, we modified the kernel side of Linux audit to perform such filtering
in order to only log files modified by the monitored services.
Then, we specified a monitoring rule that relies on such filtering.

We developed a userland daemon that listens to an audit netlink socket
and processes the events generated by our monitoring rules.
Then, by parsing them, our daemon can log which files a monitored service modified.
To that end, we create a file hierarchy under a per-service private directory.
For example, if the service \verb|abc.service| modified the file \verb|/a/b/c/test|,
we create an empty file \verb|/private/abc.service/a/b/c/test|.
This solution allows us to log modified files without keeping
a data structure in memory.

\section{Evaluation}
\label{sec:evaluation}
We performed an experimental evaluation of our approach
to answer the following questions:
\begin{enumerate}
    \item How effective are our responses at stopping malicious behaviors in case a service
    is compromised?
    \item How effective is our approach at selecting cost-sensitive responses that withstand an intrusion?
    \item What is the impact of our solution on the availability or responsiveness
    of the services?
    \item How much overhead our solution incurs on the system resources?
    \item Do services continue to function (\ie no crash)
          when they are restored with less privileges that they initially needed?
\end{enumerate}

For the experiments, we installed Fedora Server 28 with the Linux kernel 4.17.5,
and we compiled the programs with GCC 8.1.1.
We ran the experiments that used live malware
in a virtualized environment to control malware propagation
(see~\aref{appendix:virtuallab} for more details).
While malware could use anti-virtualization
techniques~\cite{paleari2009redpills,chen2008antivirt},
to the best of our knowledge, none of our samples
used such techniques.%
\footnote{This is consistent with the study of \citet{cozzi2018linuxmalware}
    that showed that in the $10\ 548$ Linux malware they studied,
    only \SI{0.24}{\percent} of them tried to detect if they were in a virtualized environment.}
We executed the rest of the experiments on bare metal
on a computer with an AMD PRO A12-8830B R7 at \SI{2.5}{\giga\hertz},
\SI{12}{\gibi\byte} of RAM,
and a \SI{128}{\giga\byte} Intel SSD 600p Series.

Throughout the experiments, we tested our implementation
on different types of services,
such as web servers (nginx~\cite{nginx}, Apache~\cite{apache_httpd}),
database (mariadb~\cite{mariadb}),
work queue (beanstalkd~\cite{beanstalkd}),
message queue (mosquitto~\cite{mosquitto}),
or git hosting services (gitea~\cite{gitea}).
It shows that our approach is applicable to a diverse set of services.

\subsection{Responses Effectiveness}
Our first experiments focus on how effective our responses against
distinct types of intrusions are.
We are not interested, per se, in the vulnerabilities that attackers can exploit,
but on how to stop attackers from performing malicious actions after they have infected
a service.
Here we do not focus on response selection,
which is discussed in~\autoref{sec:experiment:subsec:response_selection}.

The following list describes the malware and attacks used
(see~\aref{appendix:samples} for the hashes of the malware samples):
\begin{description}
    \item[Linux.BitCoinMiner]
          Cryptocurrency mining malware that connects to a mining pool using
          attackers-controlled credentials~\cite{linuxbcm}.

    \item[Linux.Rex.1]
          Malware that joins a \gls{P2P} botnet to receive instructions to
          scan systems for vulnerabilities to replicate itself,
          elevate privileges by scanning for credentials on the machine,
          participate in a \gls{DDOS} attack,
          or send spam~\cite{linuxrex}.

    \item[Hakai]
          Malware
          that receives instructions from a \gls{C2} server to launch
          \gls{DDOS} attacks, and to infect other systems by brute forcing credentials
          or exploiting vulnerabilities in routers~\cite{hakai,linuxfgt}.

    \item[Linux.Encoder.1] Encryption ransomware that encrypts files commonly
          found on Linux servers
          (\eg configuration files, or HTML files),
          and other media-related files
          (\eg JPG, or MP3), while ensuring that
          the system can boot so that the administrator can see the ransom note~\cite{linuxencoder}.

    \item[GoAhead exploit]
          Exploit that gives remote code execution to an attacker on
          all versions of the GoAhead embedded web server prior to \verb|3.6.5|~\cite{goaheadExploit}.
\end{description}

Our work does not focus on detecting intrusions but
on how to recover from and withstand them.
Hence, we selected a diverse set of malware and attacks that covered various malicious behaviors,
with different malicious behaviors.

For each experiment,
we start a vulnerable service,
we checkpoint its state,
we infect it,
and we wait for the payload to execute (\eg encrypt files).
Then, we apply our responses and
we evaluate their effectiveness.
We consider the restoration successful if the service is still functioning
and its state corresponds to the one that has been checkpointed.
Finally, we consider the responses effective if we cannot reinfect the service
or if the payload cannot achieve its goals anymore.

\begin{table}[ht]
    \centering
    \caption{Summary of the experiments that evaluate the effectiveness of the responses against various malicious behaviors}
    \label{tbl:malware_experiments}
    \resizebox{\columnwidth}{!}{
        \begin{tabular}{lll}
            \toprule
            Attack Scenario    & Malicious Behavior              & \specialcell{Per-service  \\Response Policy} \\
            \midrule
            Linux.BitCoinMiner & Mine for cryptocurrency         & Ban mining pool IPs       \\
            Linux.BitCoinMiner & Mine for cryptocurrency         & Reduce CPU quota          \\
            Linux.Rex.1        & Determine \acrshort{C2} server  & Ban bootstrapping IPs     \\
            Hakai              & Receive data from \acrshort{C2} & Ban C\&C servers' IPs     \\
            Linux.Encoder.1    & Encrypt data                    & Read-only filesystem      \\
            GoAhead exploit    & Exfiltrate via network          & Forbid connect syscall    \\
            GoAhead exploit    & Data theft                      & Render paths inaccessible \\
            \bottomrule
        \end{tabular}
    }
\end{table}

The results we obtained are summarized in~\autoref{tbl:malware_experiments}.
In each experiment, as expected, our solution successfully restored the service after
the intrusion to a previous safe state.
In addition, as expected, each response was able to withstand a reinfection
for its associated malicious behavior and only impacted the specific service
and not the rest of the system.

\subsection{Cost-Sensitive Response Selection}
\label{sec:experiment:subsec:response_selection}
Our second set of experiments focus on how effective is our approach at selecting
cost-sensitive responses.
We chose Gitea, a self-hosted Git-repository hosting service
(an open source clone of the services provided by GitHub~\cite{github}), as a
use case for a service.
We chose Gitea, because it requires a diverse set of privileges
and it shows how our approach can be applied to a complex real-world service.

In our use case, we configured Gitea
with the principle of least privileges.
It means that restrictions which corresponds to responses with a cost assigned to none
are initially applied
to the service
(\eg Gitea can only listen on port 80 and 443 or Gitea
have only access to the directories and files it needs).
Even if the service follows the best practices and is properly protected,
an intrusion can still do damages and our approach handles such cases.

We consider an intrusion that compromised our Gitea service
with the \verb|Linux.Encoder.1| ransomware.%
\footnote{In our experiments, we used an exploit for the version 1.4.0~\cite{gitea_rce}.}
When executed, it encrypts all the git repositories
and the database used by Gitea.
Hence, we previously configured the policy to set
the cost of such a malicious behavior to high,%
\footnote{One would have to assign a cost for other malicious behaviors, but for the sake of
    conciseness we only show the relevant ones.}
since it would render the site almost unusable:
$mbcost(\text{"gitea"}, \text{"compromise-data-availability"}) = \text{"high"}$.

Since our focus is not on intrusion detection, we assume that the \gls{IDS} detected
the ransomware.
This assumption is reasonable with, for example,
several techniques to detect ransomware
such as API call monitoring, file system activity monitoring,
or the use of decoy resources~\cite{kharraz2015cutting}.

In practice, however, an \gls{IDS} can generate false positives, or it can
provide non-accurate values for the likelihood of the intrusion leading
to a less adequate response.
Hence, we consider three cases to evaluate the response selection:
the \gls{IDS} detected the intrusion and accurately set the likelihood,
the \gls{IDS} detected the intrusion but not with an accurate likelihood, and
the \gls{IDS} generated a false positive.

In~\autoref{tbl:ransomware_responses}, we display a set of responses
for the ransomware that we devised based on existing strategies
to mitigate ransomware, such as CryptoDrop lockdown~\cite{cryptodrop}
or Windows controlled folder access~\cite{controlled_folders}.
None of them could have been applied proactively by the developer,
because they degrade the quality of service.
We set their respective cost for the service
and the estimated performance.

\begin{table}[ht]
    \centering
    \caption{Responses to withstand ransomware reinfection with their associated cost and performance for Gitea}
    \label{tbl:ransomware_responses}
    \resizebox{\columnwidth}{!}{
        \begin{tabular}{cllll}
            \toprule
            \# & Response                                      & Cost      & Performance \\
            \midrule
            1  & Disable open system call                      & Very High & Very High   \\
            2  & Read-only filesystem except sessions folder   & High      & Very High   \\
            3  & Paths of git repositories inaccessible        & High      & Moderate    \\
            4  & Read-only paths of all git repositories       & Moderate  & Moderate    \\
            5  & Read-only paths of important git repositories & Low       & Low         \\
            6  & Read-only filesystem                          & Critical  & Very High   \\
            \bottomrule
        \end{tabular}
    }
\end{table}

Now let us consider the three cases previously mentioned.
In the first case, the \gls{IDS} detected the intrusion and considered
the intrusion very likely.
After computing the Pareto-optimal set, we have three possible responses left
(2, 4, and 5).
The risk computed is $risk(\text{"high"}, \text{"very likely"}) = \text{high}$.
The response selection then prioritizes performance and selects the response 2 that sets the filesystem
as read-only protecting all information stored by Gitea (git repositories and its database).
The only exception is the folder used by Gitea to store sessions since having this folder read-only
would render the site unusable, thus it is a core function (see the cost critical in~\autoref{tbl:ransomware_responses}).
Gitea is restored with all the encrypted files.
The selected response prevents the attacker to reinfect the service since
the exploit require write accesses.
In terms of quality of service, users can connect to the service
and clone repositories, but due to the response
a new user cannot register and users cannot push to repositories.
Hence, this response is adequate since the service cannot get reinfected,
core functions are maintained, and other functions previously mentioned are available.

In the second case, the \gls{IDS} detected the intrusion but considered
the intrusion very unlikely while the attacker managed to infect the service.
The risk computed is $risk(\text{"high"}, \text{"very unlikely"}) = \text{low}$.
The response selection then prioritizes cost and selects the response 5 that sets
a subset of git repositories (the most important ones for the organization) as read-only.
With this response, the attacker managed to reinfect the service and the ransomware
encrypted many repositories, but not the most important ones.
In terms of quality of service, users can still access the protected repositories,
but due to the intrusion users cannot login anymore and they cannot clone the encrypted repositories
(\ie Gitea shows an error to the user).
Hence, the response is less adequate when the \gls{IDS} provides an incorrect value for the
likelihood of the intrusion, since the malware managed to encrypt many repositories,
but the core functions of Gitea are maintained.

In the third case, the \gls{IDS} detected an intrusion
with the likelihood being very unlikely, but it is in fact a false positive.
The risk computed is $risk(\text{"high"}, \text{"very unlikely"}) = \text{low}$.
It is similar to the previous case where the response selected is response 5 due to a low risk.
However, in this case, there is no actual ransomware.
In terms of quality of service, users have access to most functions
(\eg login, clone all repositories, or add issues), they just cannot push modifications
to the protected repositories.
It shows that even with false positives, our approach minimizes the impact on the quality
of service.
Once an analyst classified the alert as a false positive,
the administrator can make the service leave the degraded mode.

\subsection{Availability Cost}
In this subsection, we detail the experiments that evaluate the availability cost
for the checkpoint and restore procedures.

\subsubsection{Checkpoint}
\label{sec:evaluation:availability:checkpoint}
Each time we checkpoint a service, we freeze its processes.
As a result, a user might notice a slower responsiveness from the service.
Hence, we measured the time to checkpoint four common services:
Apache HTTP server (v2.4.33),
nginx (v1.12.1),
mariadb (v10.2.16),
and beanstalkd (v1.10).
We repeated the experiment 10 times for each service.
In average the time to checkpoint was always below
\SI{300}{\milli\second}.
\autoref{tbl:perf_checkpoint_time} of~\aref{appendix:checkpoint_restore_time}
gives more detailed timings.

The results show that our checkpointing
has a small, but acceptable availability cost.
We do not lose any connection, we only
increase the requests' latency when the service is frozen.
Since the latency increases only for a small period of time (maximum \SI{300}{\milli\second}),
we consider such a cost acceptable.
In comparison, SHELF~\cite{xiong2009shelf} incurs a \SI{7.6}{\percent} latency overhead
for Apache during the whole execution of the system.

\subsubsection{Restore}
We also evaluated the time to restore
the same services
used in~\autoref{sec:evaluation:availability:checkpoint}.
In average, it took less than
\SI{325}{\milli\second}.
\autoref{tbl:perf_restore_time} of~\aref{appendix:checkpoint_restore_time}
gives more detailed timings.

In contrast to the checkpoint, the restore procedure loses all connections due to
the kill operation.
The experiment, however, show that the time to restore a service is small
(less than \SI{325}{\milli\second}).
For example, in comparison, CRIU-MR~\cite{Webster2018criumr} took \SI{2.8}{\second} in average
to complete their restoration procedure.

\subsection{Monitoring Cost}
\label{sec:evaluation:monitoring_cost}
As detailed in~\autoref{sec:implementation:monitoring},
our solution logs the path of any file modified by a monitored service.
This monitoring, however, incurs an overhead
for every process executing on the system (\ie even for the non-monitored services).
There is also an additional overhead for monitored services that perform write accesses
due to the audit event generated by the kernel and then processed by our daemon.

Therefore, we evaluated both overhead by running
synthetic and real-world workload benchmarks,
from the Phoronix test suite~\cite{phoronix},
for three different cases:
\begin{enumerate}
    \item no monitoring is present (\textbf{baseline})
    \item monitoring rule enabled, but the service running the benchmarks \textbf{is not monitored}
          (no audit events are triggered)
    \item monitoring rule enabled and the service \textbf{is monitored} (audit events are triggered).
\end{enumerate}

\subsubsection{Synthetic Benchmarks}
We ran synthetic I/O benchmarks that stress the system by performing
many open, read, and write system calls:
\verb|compilebench|~\cite{compilebench},
\verb|fs-mark|~\cite{fsmark},
and \verb|Postmark|~\cite{katcher1997postmark}.
\verb|compilebench| emulates disk I/O operations related to the compilation of a
kernel tree, reading the tree, or its creation.
\verb|fs-mark| creates
files and directories, at a given rate and size, either synchronously or asynchronously.
\verb|Postmark| emulates an email server by performing a given number of transactions
that create, read, append to, and delete files of varying sizes.

The results of the \verb|read compiled tree| test of the \verb|compilebench| benchmark
confirmed that the overhead is only due to \verb|open| system calls with write access mode.
This test only reads files and we did not observe any noticeable overhead
(less than \SI{1}{\percent}, within the margin of error).

\begin{figure}[ht]
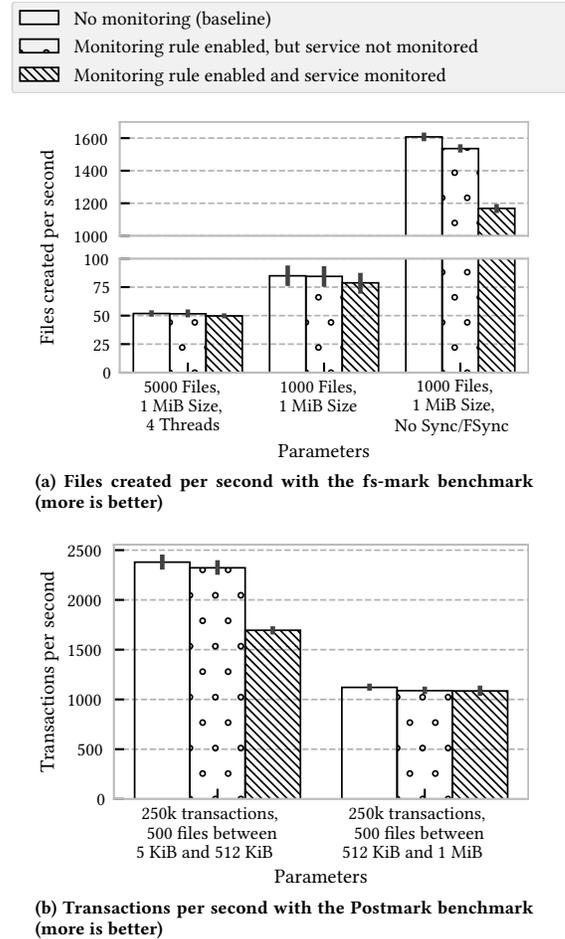

    \centering
    \captionsetup[subfigure]{width=0.78\columnwidth}
    \resizebox{0.9\columnwidth}{!}{
        \input{pgf/perf_write_monitoring_overhead_legend.tex}
    }

    \subfloat[Files created per second with the fs-mark benchmark (more is better)\label{fig:perf_fs-mark}]{
        \resizebox{0.78\columnwidth}{!}{
            \input{pgf/perf_write_monitoring_overhead_fs-mark.tex}
        }
    }

    \subfloat[Transactions per second with the Postmark benchmark (more is better)\label{fig:perf_postmark}]{
        \resizebox{0.78\columnwidth}{!}{
            \input{pgf/perf_write_monitoring_overhead_postmark.tex}
        }
    }
    \Description{There are two graphs.
    The first one shows three histograms comparing the overhead based on three cases.
    The last two histograms show us that there is a higher overhead when creating many files without synchronization.
    The second graph shows two histograms where the first histogram shows more overhead when writing many small files than the second histogram with bigger files.}
    \caption{Results of synthetic benchmarks to measure the overhead of the monitoring}
    \label{fig:perf_write_monitoring_overhead_synthetic}
\end{figure}

We now focus on the results of the \verb|fs-mark| and \verb|Postmark| benchmarks,
illustrated in~\autoref{fig:perf_write_monitoring_overhead_synthetic}.
In both experiments, we notice a small overhead when the service is not monitored
(between \SI{0.6}{\percent} and \SI{4.5}{\percent}).
With \verb|fs-mark| (\autoref{fig:perf_fs-mark}),
when writing 1000 files synchronously,
we observe a \SI{7.3}{\percent} overhead.
In comparison, when the files are written asynchronously,
there is a \SI{27.3}{\percent} overhead.
With \verb|Postmark| (\autoref{fig:perf_postmark}), we observe
a high overhead (\SI{28.7}{\percent}) when it
writes many small files (between \SI{5}{\kibi\byte} and \SI{512}{\kibi\byte})
but remains low (\SI{3.1}{\percent}) with
bigger files (between \SI{512}{\kibi\byte} and \SI{1}{\mebi\byte}).

In summary, these synthetic benchmarks show that the worst case for our monitoring
is when a monitored service writes many small files asynchronously in burst.

\subsubsection{Real-world Workload Benchmarks}
To have a different perspective than the synthetic benchmarks, we chose two benchmarks that
use real-world workloads:
\verb|build-linux-kernel| measures the time to compile the Linux kernel%
\footnote{While \verb|build-linux-kernel| is CPU bound, it also performs many system calls,
    such as opening files to store the output of the compilation.}
and \verb|unpack-linux| measures the time to extract the archive of the Linux kernel source code.

When the service is monitored, the overhead is only significant
with \verb|unpack-linux|
where we observe
a \SI{23.7}{\percent} overhead.
It concurs with our results from the synthetic benchmarks:
writing many small files asynchronously incurs a significant overhead
when the service is monitored
(the time to decompress a file in this benchmark is negligible).
With \verb|build-linux-kernel|,
we observe a small overhead  (\SI{1.1}{\percent}) even when the service is monitored
(the time to compile the source code masks the overhead of the monitoring).

In comparison, SHELF~\cite{xiong2009shelf} has a \SI{65}{\percent} overhead
when extracting the archive of the Linux kernel source code,
and an \SI{8}{\percent} overhead when building this kernel.

In conclusion, both the synthetic and non-synthetic benchmarks show that our solution
is more suitable for workloads that do not write many small files
asynchronously in burst.
For instance, our approach would be best suited to protect services
such as web, databases, or video encoding services.

\subsection{Storage Space Overhead}
Checkpointing services requires storage space to save the checkpoints.
To evaluate the disk usage overhead,
we checkpointed the same four services
used in~\autoref{sec:evaluation:availability:checkpoint}.
Each checkpoint took respectively
\SI{26.2}{\mebi\byte},
\SI{7.5}{\mebi\byte},
\SI{136.0}{\mebi\byte},
and \SI{130.1}{\kibi\byte}
of storage space.
The memory pages dumps took at least \SI{95.3}{\percent}
of the size of their checkpoint.
Hence, if a service uses more memory under load (\eg Apache),
its checkpoint would take more storage space.

\subsection{Stability of Degraded Services}
We tested our solution on
web servers (nginx, Apache),
databases (mariadb),
work queues (beanstalkd),
message queues (mosquitto),
and git hosting services (gitea).
None of the services crashed when restored with a policy that removed
privileged that they required (\ie in a degraded mode).
The reason is twofold.

First, we provided a policy that specified the responses with a critical cost.
Therefore, our solution never selected a response that removes a privilege needed
by a core function.
Second, the services checked for errors when performing various operations.
For example, if a service needed a privilege that we removed,
it tried to perform the operation and failed, but only logged an error and did not crash.
If we generalize our results, it means our solution will not make other services
(that we did not test) crash if they properly check for error cases.
This practice is common, and it is often highlighted by the compiler
when this is not the case.

\section{Discussion}
\label{sec:discussion}
Let us now discuss non-exhaustively limitations or areas that would need further
work.

\begin{description}
    \item[False Positives]
          Our approach relies on an \gls{IDS}, hence we inherit its limitations.
          In the case of false positives, the recovery and response procedures would impact
          the service availability, despite thwarting no threat.
          Our approach, however, minimizes this risk by considering the likelihood
          of the intrusion for the selection of cost-sensitive responses
          and by ensuring that we maintain core functions.

    \item[CRIU Limitations]
          At the moment, CRIU cannot support all applications, since
          it has issues when handling external resources or graphical applications.
          For example, if a process has opened a device to have direct access to some hardware,
          checkpointing its state may be impossible
          (except for virtual devices not corresponding to any physical devices).
          Since our implementation relies on CRIU, we inherit its limitations.
          Therefore, at the moment, our implementation is better suited for system services
          that do not have a graphical part and do not require direct access to some hardware.

    \item[Service Dependencies]
          In our work, at the moment,
          we only use the service dependency graph provided by the service manager
          (\eg systemd)
          to recover and checkpoint dependent services together.
          We could also use this same graph to provide more precise response selection
          by taking into account the dependency between services,
          their relative importance, and to propagate the impact a malicious behavior can have.
          It could be used as a weight (in addition to the risk) to select optimal responses.
          Similar, but network-based, approaches have been heavily studied in
          the past~\cite{toth2002evaluating,kheir2009dependency,shameli2018dynamic}.

    \item[Models Input]
          For our cost-sensitive response selection, we first need to associate an intrusion
          to a set of malicious behaviors, and the course of action to stop
          these behaviors.
          While standards exist to share threat information~\cite{stix} and
          malicious behaviors~\cite{maec,EMA,maec_hierarchy}
          exhibited by malware, or attackers in general,
          we were not able to find open sources that provided them directly
          for the samples we used.
          This issue might be related to the fact that, to the best of our knowledge,
          no industry solution would exploit such information.
          In our experiments, we extracted information about malicious behaviors
          from textual descriptions~\cite{linuxbcm,linuxencoder,linuxfgt,linuxrex,hakai}
          and reused the existing standards to describe such malicious
          behaviors~\cite{maec,EMA,maec_hierarchy}.
          Likewise, we extracted information about responses to counter
          such malicious
          behaviors from textual descriptions~\cite{linuxbcm,linuxencoder,linuxfgt,linuxrex,hakai,cryptodrop,controlled_folders}.

    \item[Generic Responses]
          If we do not have precise information about the intrusion, but only a generic
          behavior or category associated to it (one of the top elements in the malicious behaviors
          hierarchy), we can automatically consider generic responses.
          For example, with ransomware we know that responses that either render the filesystem read-only
          or only specific directories will work.
          We would not know that, for instance, blocking a specific system call would have stopped
          the malware, but we know what all ransomware need, and we can respond accordingly.
          Such generic responses might help mitigate the lack of precise information.
\end{description}
\section{Conclusion and Future Work}
\label{sec:conclusion}
This work provides an intrusion survivability approach for commodity \glspl{OS}.
In contrast to other intrusion recovery approaches, our solution
is not limited to restoring files or processes, but it also applies
responses to withstand a potential reinfection.
Such responses enforce per-service privilege restrictions and resource quotas
to ensure that the rest of the system is not directly impacted.
In addition, we only restore the files modified by the infected service
to limit the restoration time.
We devised a framework to select cost-sensitive responses
that do not disable core functions of services.
We specified the requirements for our approach and proposed an architecture satisfying its requirements.
Finally, we developed and evaluated a prototype for Linux-based systems by modifying
systemd, Linux audit, CRIU, and the Linux kernel.
Our results show that our prototype withstands known Linux attacks.
Our prototype only induces a small overhead, except with
I/O-intensive services that create many small files asynchronously in burst.

In the future,
we would like to investigate how we could automatically adapt the system
to gradually remove the responses
that we applied to withstand a reinfection.
Such a process involves being able to automatically fix
the vulnerabilities
or to render them non-exploitable.

\appendix
\section*{Acknowledgments}
The authors would like to thank
Pierre Belgarric and
Maugan Villatel
for their helpful comments, feedback, and proofing of earlier versions of this paper.
We would also like to thank
Stuart Lees,
Daniel Ellam, and
Jonathan Griffin,
for their help in setting up and running some of the experiments using their isolated and virtualized environment.
In addition, we would like to thank the anonymous reviewers for their feedback.
Finally, we would like to thank Hybrid Analysis for providing access
to malware samples used in our experiments.

\section{Examples of Models}
\label{appendix:risk_matrix}
\label{appendix:models}

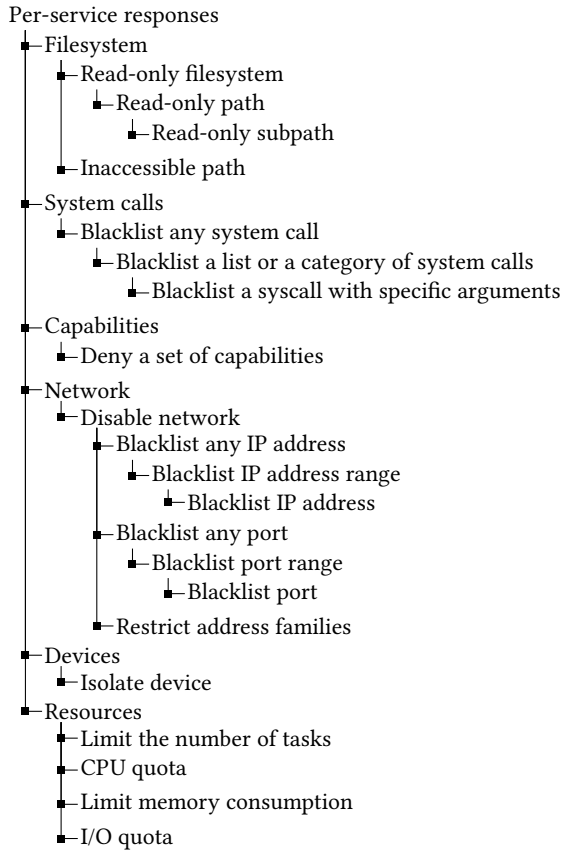
\begin{figure}[ht]
    \begin{forest}
        for tree={
            inner sep=0.1em, s sep=0.3em,
            font=\normalfont,
            grow'=0,
            child anchor=west,
            parent anchor=south,
            anchor=west,
            calign=first,
            edge path={
                \noexpand\path [draw, \forestoption{edge}]
                (!u.south west) +(0.8em,0) |- node[fill,inner sep=0.15em] {} (.child anchor)\forestoption{edge label};
            },
            before typesetting nodes={
                if n=1
                {insert before={[,phantom]}}
                {}
            },
            fit=band,
            before computing xy={l=1.5em},
        },
        [Per-service responses
            [Filesystem
                [Read-only filesystem
                [Read-only path
                    [Read-only subpath]
                ]
                ]
                [Inaccessible path]
            ]
            [System calls
                [Blacklist any system call
                    [Blacklist a list or a category of system calls
                        [Blacklist a syscall with specific arguments]
                    ]
                ]
            ]
            [Capabilities
                [Deny a set of capabilities]
            ]
            [Network
                [Disable network
                    [Blacklist any IP address
                        [Blacklist IP address range
                            [Blacklist IP address]
                        ]
                    ]
                    [Blacklist any port
                        [Blacklist port range
                            [Blacklist port]
                        ]
                    ]
                    [Restrict address families]
                ]
            ]
            [Devices
                [Isolate device]
            ]
            [Resources
                [Limit the number of tasks]
                [CPU quota]
                [Limit memory consumption]
                [I/O quota]
            ]
        ]
    \end{forest}
    \Description{A hierarchy of per-service responses including, for example, the filesystem at a top-level category with the "read-only filesystem" response and "read-only path" as a child.}
    \caption{Example of a non-exhaustive per-service response hierarchy}
    \label{fig:response_hierarchy_example}
\end{figure}

\autoref{fig:response_hierarchy_example} is an example of a non-exhaustive per-service
response hierarchy that can
be used for the hierarchy defined in~\autoref{sec:policies:malicious_behaviors_and_responses}.
Note that for each response with arguments (\eg read-only paths or blacklisting IP addresses),
the hierarchy provides a sub-response with a subset of the arguments.
For example, if there is a response that puts \verb|/var| read-only, there
is also the responses that puts \verb|/var/www| read-only.
It means that if an administrator only specified the cost of putting \verb|/var| read-only,
but a response, among the set of possible responses, sets \verb|/var/www| read-only,
our response selection framework uses the cost of its parent (\ie \verb|/var|).

\autoref{tbl:risk_matrix} is an example of a risk matrix that can be used.
This matrix can vary depending on the risk attitude (risk averse, risk neutral, or risk seeking).

\definecolor{RiskLow}{HTML}{FEE8C8}
\definecolor{RiskModerate}{HTML}{FDBB84}
\definecolor{RiskHigh}{HTML}{E34A33}
\newcommand{\RiskLow}{\cellcolor{RiskLow} L}
\newcommand{\RiskModerate}{\cellcolor{RiskModerate} M}
\newcommand{\RiskHigh}{\cellcolor{RiskHigh} H}
\begin{table}[ht]
    \centering
    \caption{Example of a $5 \times 5$ risk matrix that follows the requirements for our risk assessment}
    \label{tbl:risk_matrix}
    \setlength{\tabcolsep}{3pt}
    \newcommand{\prob}[2]{\newline \hspace*{\fill}#1 -- #2\hspace*{\fill}}
    \resizebox{\columnwidth}{!}{
        \begin{tabular}{m{6em}x{4em}x{4em}x{4em}x{4em}x{4em}}
            \toprule
                                             & \multicolumn{5}{c}{Malicious Behavior Cost}                                                                 \\
            \cmidrule{2-6}
            Confidence \newline (Likelihood) & Very low                                    & Low           & Moderate      & High          & Very high     \\
            \midrule
            Very likely                      & \RiskLow                                    & \RiskModerate & \RiskHigh     & \RiskHigh     & \RiskHigh     \\
            Likely                           & \RiskLow                                    & \RiskModerate & \RiskModerate & \RiskHigh     & \RiskHigh     \\
            Probable                         & \RiskLow                                    & \RiskLow      & \RiskModerate & \RiskModerate & \RiskHigh     \\
            Unlikely                         & \RiskLow                                    & \RiskLow      & \RiskLow      & \RiskModerate & \RiskModerate \\
            Very unlikely                    & \RiskLow                                    & \RiskLow      & \RiskLow      & \RiskLow      & \RiskLow      \\
            \bottomrule
        \end{tabular}
    }
\end{table}

\section{Evaluation Details}

\subsection{Setup of the Virtualized Environment}
\label{appendix:virtuallab}
We ran the experiments regarding the effectiveness of our responses
in a virtualized environment.
It helped us control malware propagation
and their behavior in general.

The setup consisted of an isolated network connected to the Internet with multiple \glspl{VLAN},
two \glspl{VM}, and a workstation.
We executed the infected service on a \gls{VM} connected to an isolated
\gls{VLAN} with access to the Internet.
We connected the second \gls{VM}, that executes the network sniffing tool (tcpdump),
to another \gls{VLAN} with port mirroring from the first \gls{VLAN}.
Finally, the workstation, connected to another isolated \gls{VLAN},
had access to the server managing the \glspl{VM},
the \gls{VM} with the infected service, and the network traces.

\subsection{Malware Samples}
\label{appendix:samples}
\begin{table}[ht]
    \centering
    \caption{Malware used in our experiments with the SHA-256 hash of the samples}
    \label{tbl:malware_samples_hashes}
    \resizebox{\columnwidth}{!}{
    \begin{tabular}{ll}
        \toprule
        Malware            & SHA-256\\
        \midrule
        Linux.BitCoinMiner & {\footnotesize \texttt{690aea53dae908c9afa933d60f467a17ec5f72463988eb5af5956c6cb301455b}}\\
        Linux.Rex.1        & {\footnotesize\texttt{762a4f2bf5ea4ff72fce674da1adf29f0b9357be18de4cd992d79198c56bb514}}\\
        Linux.Encoder.1    & {\footnotesize\texttt{18884936d002839833a537921eb7ebdb073fa8a153bfeba587457b07b74fb3b2}}\\
        Hakai              & {\footnotesize\texttt{58a5197e1c438ca43ffc3739160fd147c445012ba14b3358caac1dc8ffff8c9f}}\\
        \bottomrule
    \end{tabular}
    }
\end{table}

In~\autoref{tbl:malware_samples_hashes}, we list the malware samples used in our experiments alongside
their respective SHA-256 hash.

\subsection{Checkpoint and Restore Operations}
\label{appendix:checkpoint_restore_time}

\begin{figure}[ht]
    \centering
    \resizebox{\columnwidth}{!}{
        \input{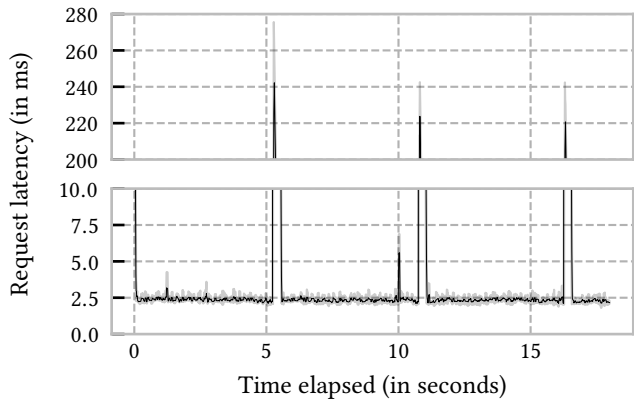}
    }
    \Description{A graph showing the impact of three checkpoints where each checkpoint incurs a noticeable increase in latency (from \SI{2}{\milli\second} to approximately \SI{240}{\milli\second} when checkpointing).}
    \caption{Impact of checkpoints on the latency of HTTP requests made to an nginx server (less is better)}
    \label{fig:perf_checkpoint_latency_nginx}
\end{figure}

In~\autoref{fig:perf_checkpoint_latency_nginx}, we illustrate the results of
the availability cost that users could perceive by measuring
the latencies of HTTP requests made to an nginx server.
We generated 100 requests per second for 20 seconds
with the HTTP load testing tool Vegeta~\cite{vegeta}.
During this time, we checkpointed nginx at approximately 5, 11, and 16 seconds.
We repeated the experiment three times.
The output gave us the latency of each request, and we applied
a moving average filter with a window size of 5.
As mentioned previously, all requests were successful (\ie no errors or timeouts)
and the maximum latency during a checkpoint was \SI{286}{\milli\second}.

In~\autoref{tbl:perf_checkpoint_time}, we show the time measured
to perform the different operations executed during a checkpoint:
initialize (\ie to initialize structures, to create or open directories, and to freeze processes),
snapshot of the filesystem,
serialization of the service's metadata,
and checkpointing of the processes using CRIU.
The time to perform this last operation varies depending on
the service (\eg the number of processes, memory used, or files opened).

\begin{table}[ht]
    \centering
    \caption{Time to perform the checkpoint operations of a service}
    \label{tbl:perf_checkpoint_time}
    \resizebox{\columnwidth}{!}{
        \begin{tabular}{lllrrr}
            \toprule
            \multicolumn{3}{l}{Checkpoint Operation} & Mean                        & \specialcell{Standard                          \\deviation} & \specialcell{Standard error\\of the mean} \\
            \midrule
            \multicolumn{3}{l}{%
                Service-independent operations}          &                             &                       &                        \\
            & Initialize                  & (\si{\micro\second})  & 643.20 & 90.75 & 14.35 \\
            & Checkpoint service metadata & (\si{\micro\second})  & 51.47  & 8.45  & 1.33  \\
            & Snapshot filesystem
            & (\si{\milli\second})        & 98.95                 & 1.38   & 2.19          \\
            \multicolumn{3}{l}{
                Checkpoint processes (CRIU)}             &                             &                       &                        \\
            & httpd                       & (\si{\milli\second})  & 199.24 & 11.05 & 3.49  \\
            & nginx                       & (\si{\milli\second})  & 51.59  & 3.99  & 1.26  \\
            & mariadb                     & (\si{\milli\second})  & 171.77 & 8.52  & 2.69  \\
            & beanstalkd                  & (\si{\milli\second})  & 16.25  & 1.37  & 0.43  \\
            \midrule
            \multicolumn{3}{l}{Total}                &                             &                       &                        \\
            & httpd                       & (\si{\milli\second})  & 298.88 &       &       \\
            & nginx                       & (\si{\milli\second})  & 151.24 &       &       \\
            & mariadb                     & (\si{\milli\second})  & 271.41 &       &       \\
            & beanstalkd                  & (\si{\milli\second})  & 115.89 &       &       \\
            \bottomrule
        \end{tabular}
    }
\end{table}

\begin{table}[ht]
    \centering
    \caption{Time to perform the restore operations of a service}
    \label{tbl:perf_restore_time}
    \resizebox{\columnwidth}{!}{
        \begin{tabular}{lllrrr}
            \toprule
            \multicolumn{3}{l}{Restore Operation} & Mean                     & \specialcell{Standard                         \\deviation} & \specialcell{Standard error\\of the mean} \\
            \midrule
            \multicolumn{3}{l}{%
                Kill processes}                       &                          &                       &                       \\
            & httpd                    & (\si{\milli\second})  & 16.39  & 2.52  & 1.13 \\
            & nginx                    & (\si{\milli\second})  & 19.24  & 3.69  & 1.65 \\
            & mariadb                  & (\si{\milli\second})  & 28.48  & 2.16  & 0.97 \\
            & beanstalkd               & (\si{\milli\second})  & 10.85  & 1.19  & 0.53 \\
            \multicolumn{3}{l}{
                Service-independent operations}       &                          &                       &                       \\
            & Initialize               & (\si{\micro\second})  & 209.40 & 32.07 & 7.17 \\
            & Compare Snapshots
            & (\si{\milli\second})     & 148.23                & 32.01  & 7.16         \\
            & Restore service
            metadata                              & (\si{\micro\second})     & 212.75                & 36.23  & 8.10         \\
            \multicolumn{3}{l}{
                Restore processes (CRIU)}             &                          &                       &                       \\
            & httpd                    & (\si{\milli\second})  & 132.42 & 6.09  & 2.72 \\
            & nginx                    & (\si{\milli\second})  & 59.88  & 4.88  & 2.18 \\
            & mariadb                  & (\si{\milli\second})  & 147.07 & 2.59  & 1.16 \\
            & beanstalkd               & (\si{\milli\second})  & 36.63  & 2.87  & 1.28 \\
            \midrule
            \multicolumn{3}{l}{Total}             &                          &                       &                       \\
            & httpd                    & (\si{\milli\second})  & 299.29 &       &      \\
            & nginx                    & (\si{\milli\second})  & 227.79 &       &      \\
            & mariadb                  & (\si{\milli\second})  & 324.22 &       &      \\
            & beanstalkd               & (\si{\milli\second})  & 196.16 &       &      \\
            \bottomrule
        \end{tabular}
    }
\end{table}

\bibliographystyle{ACM-Reference-Format}
    \bibliography{paper}

\end{document}